\documentclass[conference]{IEEEtran}
\IEEEoverridecommandlockouts
\usepackage{cite}
\usepackage{amsmath,amssymb,amsfonts}
\usepackage{algorithmic}
\usepackage{graphicx}
\usepackage{textcomp}
\usepackage{hyperref}
\usepackage[normalem]{ulem}
\usepackage[table,xcdraw]{xcolor}
\def\BibTeX{{\rm B\kern-.05em{\sc i\kern-.025em b}\kern-.08em
    T\kern-.1667em\lower.7ex\hbox{E}\kern-.125emX}}

\newcommand{\reviwer}[1]{\textcolor{black}{#1}}
\begin{document}

\title{LoRa Multi-Hop Networks for Monitoring Underground Mining Environments\\

\thanks{The present work has received funding from the European Union’s Horizon 2020 Marie Skłodowska Curie Innovative Training Network Greenedge (GA.No.  953775) and is co-funded by the Spanish Ministry of Science, Innovation and Universities under RF-VOLUTION project (PID2021-122247OB-I00), the SGR 00174 2021 grant.}
}

\author{
\IEEEauthorblockN{Luca Scalambrin}
\IEEEauthorblockA{
\textit{Worldsensing \& Universitat Oberta de Catalunya}\\
Barcelona, Spain \\
lscalambrin@worldsensing.com}
\and
\IEEEauthorblockN{Andrea Zanella}
\IEEEauthorblockA{
\textit{University of Padova}\\
Padova, Italy \\
andrea.zanella@unipd.it }
\and
\IEEEauthorblockN{Xavier Vilajosana}
\IEEEauthorblockA{
\textit{Universitat Oberta de Catalunya}\\
Barcelona, Spain\\
xvilajosana@uoc.edu}
}

\maketitle

\begin{abstract}
Internet of Things applications have gained widespread recognition for their efficacy in typical scenarios, such as smart cities and smart healthcare. Nonetheless, there exist numerous unconventional situations where IoT technologies have not yet been massively applied, though they can be extremely useful. One of such domains is the underground mining sector, where enhancing automation monitoring through wireless communications is of essential significance. In this paper, we focus on the development, implementation, and evaluation of a LoRa-based multi-hop network tailored specifically for monitoring underground mining environments, where data traffic is sporadic, but energy efficiency is of paramount importance. We hence define a synchronization framework that makes it possible for the nodes to sleep for most of the time, waking up only when they need to exchange traffic. Notably, our network achieves a sub \textbf{$\boldsymbol{40}$~$\boldsymbol{\mu}$}s proven synchronization accuracy between parent-child pairs with minimum overhead for diverse topologies, rendering it highly viable for subterranean operations. Furthermore, for proper network dimensioning, we model the interplay between network's throughput, frame size, and sampling periods of potential applications. Moreover, we propose a model to estimate devices' duty cycle based on their position within the multi-hop network, along with empirical observations for its validation. The proposed models make it possible to optimize the network's performance to meet the specific demands that can arise from the different subterranean use cases, in which robustness, low power operation, and compliance with radio-frequency regulations are key requirements that must be met. 
\end{abstract}



\section{Introduction}

In recent years, the Internet of Things (IoT) has witnessed remarkable growth and impact across conventional domains, such as smart agriculture, smart cities and smart healthcare \cite{iot_impact}. However, there are numerous unexplored applications where IoT could make a significant positive contribution, as is the case with monitoring in the underground mining sector \cite{underground_chapter}. This sector poses challenging working conditions for laborers, and the automation of monitoring processes could not only accelerate on-site measurements\reviwer{, including those of air quality, pressure, temperature, and structural vibrations}, but also considerably reduce the risks faced by workers, which have resulted in numerous fatalities over time \cite{underground_disaster}. One of the primary challenges in implementing IoT-based monitoring in such scenarios is the communication problem, as they predominantly consist of rock wall tunnels, where radio propagation is extremely challenging due to high signal loss \cite{propagation_underground}. Consequently, monitoring devices must exhibit reliable communication capabilities, and they should be easily deployable, considering that the locations can be difficult to access. 

Among the various IoT solutions available, LoRaWAN stands out as one of the most commonly used for monitoring rugged environments with low data rate requirements, as is the case for underground domains. This is mainly due to its robust long-range coverage, achieved through a spread spectrum technique denominated chirp spread spectrum (CSS), where the signal is modulated by chirp pulses \cite{lora_limits_2}. Nevertheless, LoRaWAN's star topology architecture, in which sensing nodes wirelessly connect to a central point known as \textit{Gateway}, presents significant limitations in subterranean galleries, primarily because a single hop is insufficient to cover the long distances that can be found in such scenarios. To overcome this challenge, a possible solution is to implement a multi-hop network, a concept extensively utilized in the past with the advent of technologies such as WirelessHART and 802.15.4-6TISCH \cite{wireless_hart_2, 6tisch}. However, these technologies were not engineered to face the challenges of harsh underground environments, and they were optimized for higher capacities than those provided by LoRa. As a consequence, fitting all the required signaling within LoRa data rates while remaining compliant with duty cycle regulations becomes unfeasible, which leads to the need for the development of a new LoRa-based multi-hop framework.

Building an energy-efficient multi-hop structure requires coordination between devices so communication processes can occur in a synchronized manner. This coordination is usually rooted in clock synchronization approaches that ensure a common notion of time among the entire network \cite{mesh_sync}. A usual method to  achieve synchronization is to rely on a network protocol in which parent nodes act as time reference for their children, using data and/or control  traffic to minimize the clock drift between peers \cite{drift}. However, maintaining network synchronization often comes at the expense of sending extra packets, which can negatively impact the network. Consequently, the LoRa multi-hop protocol must keep overhead to a minimum, so as to run efficiently on extremely constrained devices whilst adhering to regional regulations.

In this work, we propose a Time-Division Multiple Access (TDMA) scheme explicitly designed to support multi-hop LoRa-based communication in tree-shaped wireless networks. The proposed protocol has been implemented and tested on real devices to prove its effectiveness and practicality. To summarize, the primary contributions of this work are: \\1) Design, implementation and testing of a TDMA multi-hop protocol for LoRa-based wireless networks, with a sub $40$~$\mu$s father-child synchronization accuracy; \\2) Theoretical analysis of the inter-dependencies of frame size, throughput and sampling period in a multi-hop network;  \\3) Model to estimate devices duty cycle depending on their position within the multi-hop network, jointly with measurements for validation.

All these contributions together aim to provide a wider perspective on the feasibility of LoRa multi-hop for industrial scenarios, considering the limiting factors for its practical adoption, such as the band regulations. As far as we know, this is the first study that includes such a perspective. The remainder of this paper is organized as follows: the related work is briefly commented on Sec.~\ref{sec:related_work}. In Sec.~\ref{sec:tdma_design} we detailed the TDMA architecture designed together with a theoretical analysis of the more important parameters, whilst the experimental setup is presented in Sec.~\ref{sec:experimental_setup} and the results are provided in Sec.~\ref{sec:results}. Finally, conclusions and future research lines are discussed in Sec.~\ref{sec:conclusions}.

\section{Related Work}
\label{sec:related_work}
In recent years, there has been a growing interest in the design and implementation of LoRa multi-hop networks, due to their high versatility in monitoring applications. In \cite{2018_lee}, the authors proposed a LoRa multi-hop architecture where the gateway is in charge of sending queries to the child nodes when a sampling operation is required. A significant limitation of this asynchronous design is that end nodes must be kept in a reception state continuously, making it unsuitable for battery-powered devices. Subsequently, Abrardo \textit{et al.} \cite{2019_abrardo} proposed an extension of LoRaWAN architecture for monitoring underground infrastructures in Italy. However, the solution only provides a line topology, which is not representative of the most common real-world underground scenarios, and there is also a lack of experimentation, as all the reported results were obtained entirely from simulations. In \cite{2019_ebi_underground}, Ebi \textit{et al.} proposed an interesting synchronous mechanism for monitoring urban drainage systems centered on a relay node that collects data from the other devices. However, the relay requires some special hardware, \textit{i.e.}, a double front-end capable of handling both the LoRaWAN layer and the mesh protocol, impacting directly on the device's cost. In addition, the  achieved synchronization accuracy is not optimal for the hardware employed, leading to longer reception times that require more energy. 

A recent study by Mugerwa \textit{et al.} \cite{2023_implicit} addresses the challenge of packet loss in LoRaWAN networks for devices that are distant from the gateway. Their approach is similar to the one proposed by the LoRa-Alliance, in which devices closer to the gateway take on the role of relays. However, a notable limitation of the method is that it only provides an additional hop, which proves insufficient for underground mining environments. In \cite{2022_mesh_library}, a library that facilitates the integration of LoRa-enabled devices into a mesh network with a routing protocol of distance-vector type is presented. Unfortunately, this library is exclusively suitable for end devices that can be held in the listening state indefinitely, \reviwer{as it is also the case of the work carried out in \cite{2020_detonation, 2020_HBEE_routing}}, rendering it impractical for most underground monitoring applications.

\begin{figure}[htpb]
    \begin{center}
        \includegraphics[keepaspectratio, scale=.38]{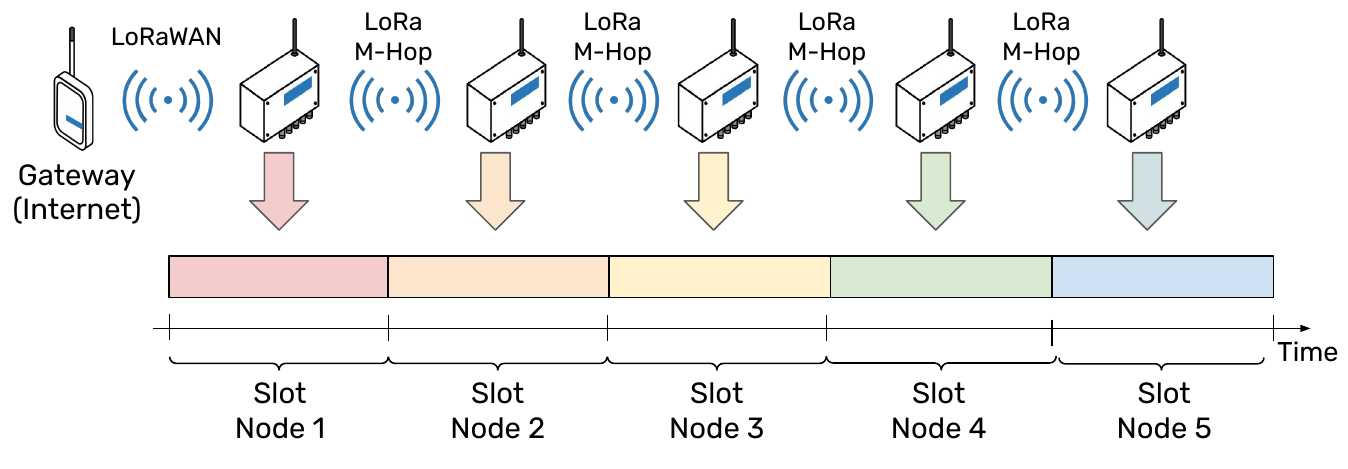}
        \caption{Time dimension division technique employed between the different devices in the network. This ensures that nodes of the system transmit and receive only in their corresponding slots.}
        \label{fig:tdma_scheme}
    \end{center}
\end{figure}

Given the inadequacy of the mentioned approaches for monitoring in underground scenarios, this study proposes a TDMA protocol specifically designed for battery-powered LoRa-enabled devices to be deployed in underground tunnels/galleries, where only one permanent-power sourced gateway is needed at the periphery of the network, as exemplified in Fig.~\ref{fig:tdma_scheme}.
\vspace{-.4cm}
\section{TDMA Architecture}
\label{sec:tdma_design}
In this section, we provide a detailed explanation of the TDMA protocol designed, together with a theoretical analysis between frame size and throughput, which is fundamental to correctly dimensioning the system. Then, we provide a model that aims to estimate the duty cycle of devices within the multi-hop network based on their position.

\subsection{Principle of Operation}
The multi-hop LoRa protocol we have developed consists of a TDMA system built at the Medium Access Control (MAC) layer on top of the LoRa physical layer (PHY). Fig.~\ref{fig:tdma_scheme} illustrates the fundamental operational principle, which consists of a tree architecture composed of one relay node and many other regular nodes. The protocol allocates the time dimension, dividing it into distinct time slots, each designated for specific nodes within the network. To establish seamless wireless connections, every node within the network executes the LoRa multi-hop protocol, ensuring successful communication with both its parent and child nodes, if any. Conversely, the relay node, being the unique gateway-reachable device, must also implement the LoRaWAN stack. Therefore, the relay node undertakes the vital role of receiving all data packets transmitted within the network under the multi-hop system and subsequently uploading them to the Gateway through LoRaWAN communication.

The multi-hop LoRa MAC layer employs a versatile framework consisting of five packet types, each encapsulated within the LoRa PHY layer. Among these packet types, \texttt{JoinRequest} and \texttt{JoinAccept} serve the purpose of enabling devices to join the network. The former is utilized by a joining node to indicate its chosen parent node, while the latter is dispatched by the relay node containing the assigned time slot number within the TDMA framework, upon a successful joining procedure. The next two packets, named \texttt{UpData} and \texttt{DownData}, correspond to the data packets used for sending application information in both the uplink and downlink directions, which take place only once per frame for each device.

\begin{figure}[htpb]
    \begin{center}
        \includegraphics[keepaspectratio, scale=.6]{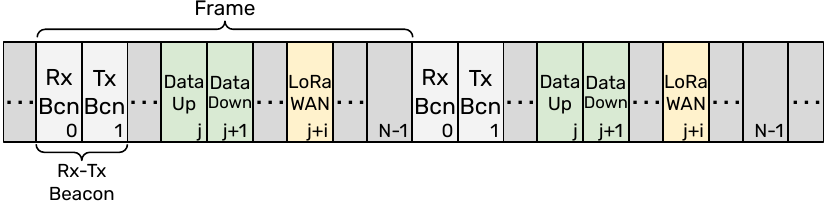}
        \caption{Frames are composed of N slots. To keep synchronization, each node receives and transmits one beacon at the beginning of the frame.}
        \label{fig:frame_scheme}
    \end{center}
\end{figure}

As it was mentioned previously, the notion of time is of paramount importance for every device in the system. In this regard, each device is equipped with a clock oscillator, whose oscillation frequency can be customized to get a certain \textit{Tick} time. Ideally, the tick duration should be the same for all devices, but real oscillators do not behave perfectly due to internal and external factors, preventing all the network's nodes from achieving the same tick period. To overcome this challenge, the multi-hop LoRa system adopts a beacon-based technique that periodically corrects the reference time across the network elements. As exemplified in Fig.~\ref{fig:frame_scheme}, the TDMA framework defines a frame period composed of $N$ slots, which are assigned to specific nodes for data or beacon exchange. Each of the mentioned frames begins with the reception of the fifth packet type, denoted as \texttt{Beacon}, which provides the reference to re-synchronize clocks. This packet contains essential information such as \texttt{NetworkID} and \texttt{SenderID}. It can also be noticed that one dedicated slot within each frame is reserved for the LoRaWAN link, which is exclusively utilized by the relay node to transmit the collected data to the Gateway. Consequently, the throughput of the entire network is limited to one LoRaWAN packet per frame.
\vspace{-.4cm}
\vspace{-.5cm}

\begin{figure}[htpb]
    \begin{center}
        \includegraphics[keepaspectratio, scale=.34]{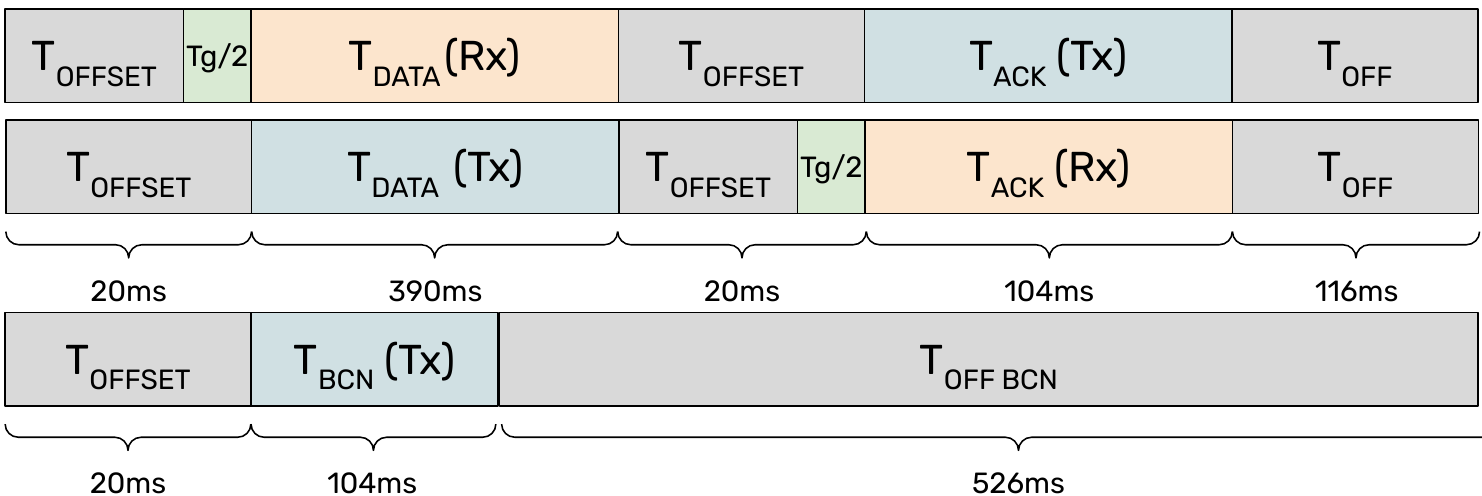}
        \caption{Slot structure for the different types of slots used for receiving (top) or transmitting (middle) a data packet, or transmitting a beacon (bottom) in the TDMA multi-hop architecture.}
        \label{fig:slot_structure}
    \end{center}
\end{figure}

Simultaneously, each slot within the frame consists of various distinct time elements, as depicted in Fig.~\ref{fig:slot_structure}, categorized based on the actions to be executed.\footnote{ Fig.~\ref{fig:slot_structure} reports the parameter values used in this study, but the framework can be adapted to other design choices.} When transmitting a data packet, devices initiate the transmission and then await the corresponding acknowledgment (\texttt{ACK}) packet. Conversely, during packet reception, devices first handle the reception process and subsequently transmit the corresponding acknowledgment message. $T_{offset}$ is employed as a buffer to start the reception or transmission, as radio chips always require extra time to reach the \texttt{ReadyToReceive} state. A slot $T_{data}$ ($T_x$ and $R_x$)  enables nodes to transmit or receive data packets of a predefined maximum size (\mbox{64 bytes} in this study) through the PHY LoRa layer, with all devices operating under the multi-hop protocol with the same spreading factor (set to $SF=9$ in our case).\footnote{The framework can potentially be adjusted to host links of different capacities, but this would increase the framework complexity and, in turn, the nodes' energy consumption. Given the typically light traffic demand of the target applications, simplicity and energy efficiency are preferred to capacity.} $T_{bcn}$ and $T_{ack}$ represent beacon and acknowledgement message duration, and they are kept as compact as possible with the goal of minimizing radio usage. 
\vspace{-.4cm}
\vspace{-.5cm}

\begin{figure}[htpb]
    \begin{center}
        \includegraphics[keepaspectratio, scale=.45]{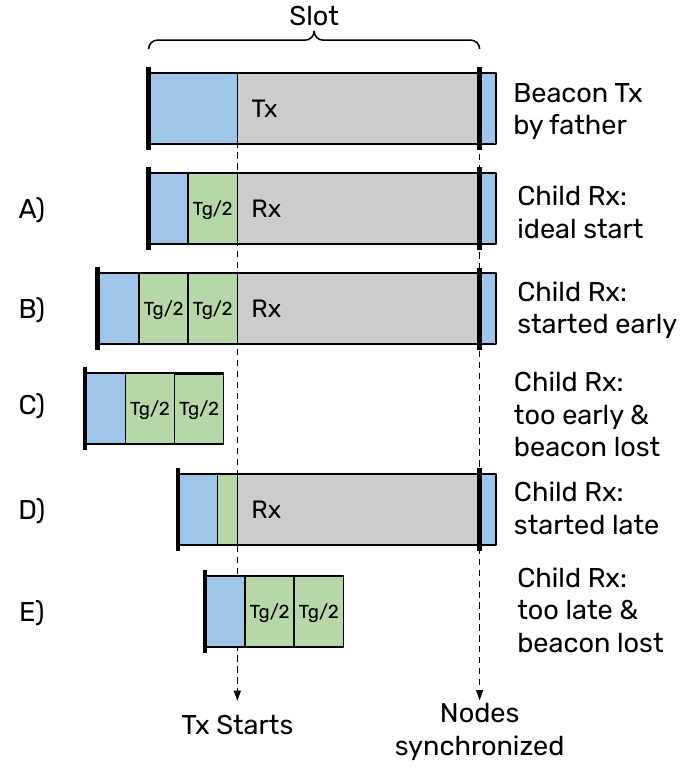}
        \caption{Schemes \textit{A} to \textit{E} exemplify the various potential states of a child node while receiving a beacon transmitted by its parent node. In cases \textit{A}, \textit{B}, and \textit{D}, the devices achieve complete synchronization, whereas synchronization does not occur in cases \textit{C} and \textit{E}.}
        \label{fig:slot_correction}
    \end{center}
\end{figure}

The final element contained within the slot is referred to as \textit{Guard Time}, $T_g$, which represents an extension of the reception window necessary in TDMA systems to compensate for clock imperfections. If $T_{F}$ denotes the frame time, which is  the time between two consecutive synchronization events, and $D_R$ the relative drift between father and child, $T_g$ window must satisfy
\begin{equation}
    T_g/2 \geq D_R T_F
    \label{eq:Tg}
\end{equation}
in order to keep the framework working at all times. It is noticeable that devices with a larger $D_R$ will require a larger $T_g$, which yields lower energy efficiency. To illustrate the most typical synchronization  cases, in Fig.~\ref{fig:slot_correction} we depict different scenarios that can occur when a node is receiving a beacon signal from its parent. The ideal case \textit{A} occurs if no drift is present between the devices, whilst \textit{B} and \textit{D} illustrate the early and late cases in which the child  is still able to get the synchronization signal, with \textit{B} being the worst case scenario in terms of energy, as the node needs to be in $R_x$ state for the entire $T_g$ interval before the actual reception starts. On the other hand, \textit{C} is too early and \textit{E} is too late to get the beacon.
\vspace{-.4cm}
\subsection{Frame Size and Network Throughput}
Once the TDMA framework is established, critical parameters, such as the number of slots $N$ within a frame or the application period, must be carefully selected in order to dimension the network properly. To analyze their impact in terms of energy consumption, it is required to consider the different power consumption states of the nodes, which heavily depend on the hardware employed. Generally, nodes with radio chips can operate in three main states: \textit{Transmission State}, \textit{Reception State}, and \textit{Sleep State}. 
Note that, we do not distinguish between the power state of the micro-controller ($\mu$C) and that of the radio interface because, with the boards considered in this study, the $\mu$C is activated any time the radio transceiver leaves the sleep state. Therefore, we classify the state of the whole node, and we denote by $P_s$, $P_{Tx}$ and $P_{Rx}$ the overall power consumption in Sleep, Transmission and Reception states, respectively. 
Additionally, we account for the energy consumed by an application running on top of the TDMA system, indicating by $P_{app}$ its power consumption and by $\tau_{app}$ its execution time. We assume the application is executed once every $k$ frames, where $k$ is an integer parameter. If $T_{SL}$ denotes the slot duration, the application period $T_{app}$ can hence be expressed as
\begin{equation}
    \label{eq:kn}
    T_{app} = k T_{F} = k T_{SL}N,
\end{equation}
where $T_{SL}$ depends on the maximum number of transmitted bytes per slot and the selected spreading factor $SF$. As both $T_{app}$ and $T_{SL}$ remain constant during execution, their product $kN$ is also  constant (the longer the frame, the lower the number of frames between two application executions). 

The mean power $P_{tot}$ absorbed by a device that receives and sends a beacon, executes the application, and enters the sleep state during the rest of the time, can be written as:
\begin{equation}
\begin{split}
    \label{eq:pow_tdma}
    P_{tot} = P_s + (P_{Rx} + P_{Tx} - 2 P_{s}) \frac{T_{bcn}}{T_{SL} N} + \\ 
    (P_{Rx}-P_{s}) \frac{T_g}{T_{SL} N} + (P_{app} - P_s) \frac{\tau_{app}}{T_{app}}.
\end{split}
\end{equation}
Recalling \eqref{eq:Tg}, the guard time is also proportional to the frame duration. Replacing $T_g$ in \eqref{eq:pow_tdma} with its lower bound $2D_RT_F$, we get 
\begin{equation}
\begin{split}
  \label{eq:pow_tdma2}
    P_{tot} = P_s + (P_{Rx} + P_{Tx} - 2 P_{s}) \frac{T_{bcn}}{T_{SL} N} + \\ 
    (P_{Rx}-P_{s}) 2D_R + (P_{app} - P_s) \frac{\tau_{app}}{k T_{SL}N}.
\end{split}
\end{equation}
From \eqref{eq:pow_tdma2}, it is evident that, to minimize power consumption, $N$ should be increased and, since $kN$ must remain constant, $k$ should be set to $1$. 

This result suggests prolonging the frame duration as much as possible, such that $T_{app} = T_F = NT_{SL}$, allowing the application to be executed once per frame. 
However, as mentioned earlier, the TDMA architecture can handle only one LoRaWAN transmission from the relay node to the Gateway per frame. Therefore, if the entire network generates more than one packet to be sent to the Gateway within a single frame, it will result in an overload situation. Hence, if $n$ represents the number of devices within the network, $T_{app}$ is limited by
\begin{equation}
    \label{eq:sampling_limit}
    n T_F \leq T_{app},
\end{equation}
which, by~\eqref{eq:kn}, can be  expressed as $n \leq k$. These expressions signify that reducing the parameter $k$, \textit{i.e.}, keeping the frame as long as $T_{app}$, yields better energy efficiency, but at the cost of imposing a limitation on the maximum number of devices that the network can accommodate without collapsing. In a practical setting, then, the frame size is limited by  $T_{app}/n$. Alternatively, it is possible to allocate more LoRaWAN slots in the frame, slightly changing the framework.

\subsection{Duty Cycle}
The duty cycle is another critical parameter that restricts radio usage in sub-GHz unlicensed bands. In Europe, for instance, ETSI established a maximum duty cycle of 1\% per channel per device. In multi-hop networks, nodes closer to the relay tend to be the root of larger sub-trees, resulting in increased data transmissions. Consequently, duty cycle restrictions become significant for these devices within dense networks. 

If we denote the number of available (orthogonal) radio channels by $c$ and the application period by $T_{app}$, and $m_i$ counts for the number of devices within the sub-tree rooted by the $i$-th device, its duty cycle $D_{C_i}$ is given by
\begin{equation}
\label{eq:dc_vs_mi}
    D_{C_i} = \frac{m_iT_{ack} + (1+m_i)T_{data} + kT_{bcn}}{T_{app}}\frac{1}{c}\,.
\end{equation}
From this expression, the duty cycle of different devices within the multi-hop network can be easily estimated. Additionally, as in some cases the duty cycle $D_{C}$ restriction can be the main constraint for $T_{app}$, the minimum application period will be given by the most restrictive condition between \eqref{eq:sampling_limit} and
\begin{equation*}
\label{eq:tapp_vs_mi}
    T_{app} \geq \frac{m_iT_{ack} + (1+m_i)T_{data} + kT_{bcn}}{D_C}\frac{1}{c}\,.
\end{equation*} 

\section{Experimental setup}
\label{sec:experimental_setup}

\begin{figure}
    \begin{center}
        \includegraphics[keepaspectratio, scale=.42]{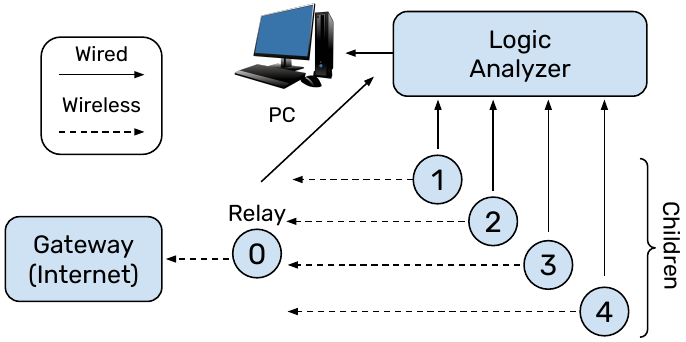}
        \caption{Setup of one relay and four children communicated via LoRa multi-hop, while a logic analyzer is wired connected to every device to facilitate GPIO signal measurements.} 
        \label{fig:setup}
    \end{center}
\end{figure}

The system measurements were conducted using a Logic Analyzer (Digilent Digital Discovery) jointly with a computer and \reviwer{an indoor setup} of LoRa-based IoT devices, as illustrated in Fig.~\ref{fig:setup}, \reviwer{where attenuators of $20$~dB were included on each device, \textit{i.e.}, $40$~dB attenuation for each link, to simulate longer distances, given the challenges of accessing underground assets for real-world experimentation}. Apart from the radio link between the IoT devices, a wired connection between the nodes and the Logic Analyzer was employed for the measurement of General Purpose Input/Output (GPIO) pin signals.
The hardware components utilized in the LoRa devices consisted of an SX1276 Radio chip and a \text{32-bit} ARM architecture micro-controller running \texttt{FreeRTOS} operating system. For the TDMA system, a 32~KHz crystal was utilized as the time source, which leads to a tick duration of $30.5$~$\mu$s. For the implementation, each frame comprised 90~slots of 21281~ticks, which results in $T_F=58.5$~s when using a constant spreading factor $SF = 9$.
\vspace{-0.1cm}
\section{Results}
\label{sec:results}
\vspace{-0.1cm}
\subsection{Synchronization error}

\begin{figure}
    \begin{center}
        \includegraphics[keepaspectratio, scale=.55]{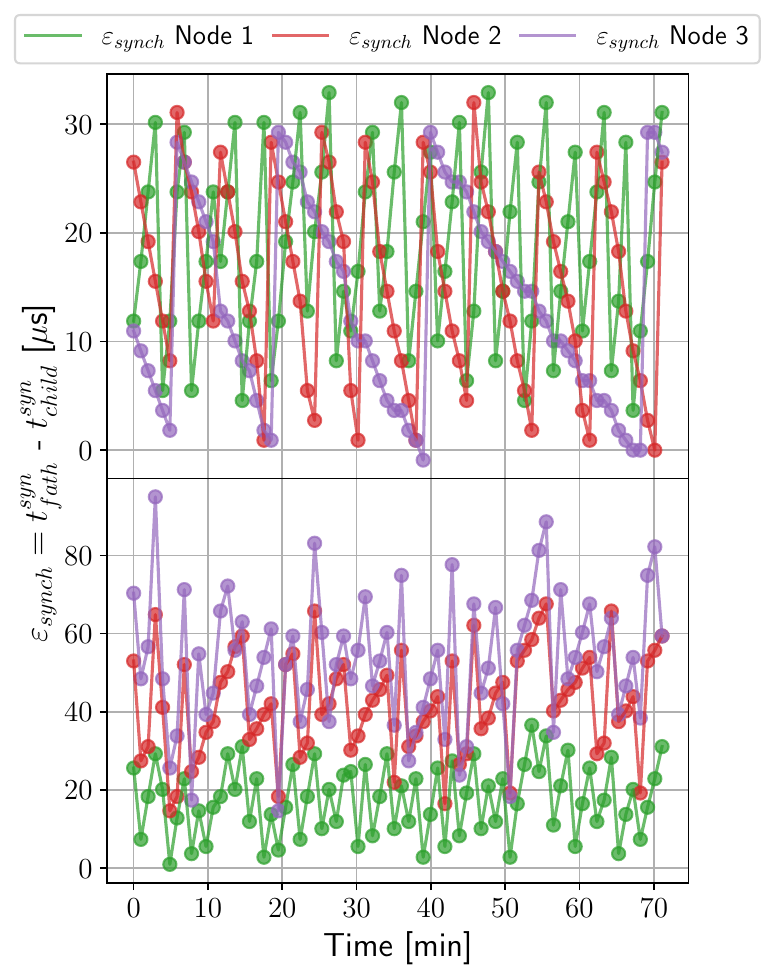}
        \caption{Synchronization error with respect to the relay (Node 0) in a star (top) and line topology (bottom).} 
        \label{fig:sync_error}
    \end{center}
\end{figure}

The synchronization error is a fundamental property of the system that quantifies the time discrepancy between two devices. To measure this parameter, both the father and children generate a signal through the GPIO pin at the end of the synchronization slot. Let $t^{syn}_{i}$ represent the time at which the synchronization event slot ends for the $i$-th device. Thus, the synchronization error, $\varepsilon_{synch}$, can be expressed as
\begin{equation*}
    \varepsilon_{synch} = t^{syn}_{fath} - t^{syn}_{child}.
\end{equation*}
Due to quantization effects, the signal generated at $t^{syn}_{i}$ can only occur at multiples of the $i$-th device's tick duration. Consequently, the maximum resolution achievable is limited to one tick duration. Fig.~\ref{fig:sync_error}~-~(top) presents the measurements of $\varepsilon_{synch}$ obtained for a \textbf{star topology} with Node 0 acting as a relay and the rest of Nodes as children, as exemplified in Fig.~\ref{fig:setup}. It is evident from the plot that the synchronization error is bounded to approximately $30.5$~$\mu$s, which corresponds to the tick duration of the clock used. As a result, this implementation represents the first LoRa multi-hop system with the maximum resolution achievable for a $32$~KHz clock.
Contrary to the star topology, Fig.~\ref{fig:sync_error}~-~(bottom) depicts the $\varepsilon_{synch}$ measurements obtained with devices arranged in a \textbf{line topology} as shown on Fig.~\ref{fig:tdma_scheme}. It must be noticed that the error is measured with respect to the relay, causing an error increment towards the end of the chain. Nevertheless, as it can be recognized, the father-child error of the node pairs $0-1$, $1-2$ and $2-3$ is always bounded to approximately $30.5$~$\mu$s, 
\reviwer{proving that every node in the network will be able to communicate accurately the relevant data generated by diverse monitoring applications.}
\vspace{-0.1cm}

\subsection{Duty Cycle}

\begin{figure*}
    \begin{center}
        \includegraphics[keepaspectratio, width=\textwidth]{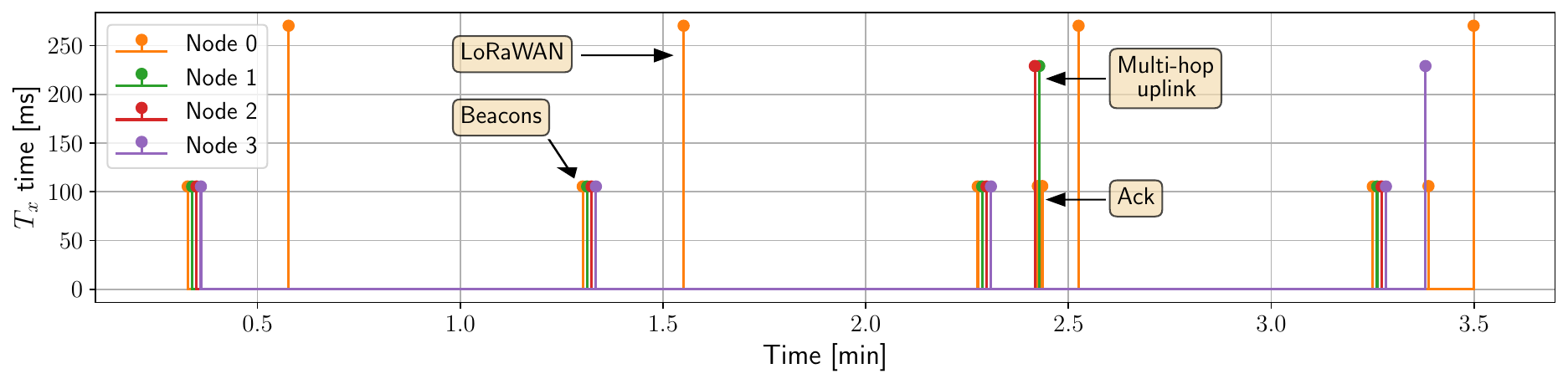}
        \caption{Transmission time of devices connected in a star topology with $T_{app} = 4T_F$, $T_F=58.5$~s and $SF=9$. Node 0 behaves as the relay of the network, while nodes 1 to 3 take on the role of children.} 
        \label{fig:tx_on_radio}
    \end{center}
\end{figure*}

For duty cycle measurements, the network was configured with a star topology as displayed in Fig.~\ref{fig:setup}, with Node 0 acting as the relay. Fig.~\ref{fig:tx_on_radio} illustrates the results obtained with a sampling period of $T_{app} = 4 T_F$ and considering only one channel, \textit{i.e.}, $c=1$. The beginning of each frame is easily identifiable due to the four transmissions of beacons, whose Time-On-Air results in $ToA = 103.4$~ms. The uplink data size used in the multi-hop network was $24$~bytes, and so the $ToA$ becomes $226.3$~ms, which can be identified in the figure together with the acknowledgment messages sent. Additionally, each frame includes a LoRaWAN transmission by the relay, which can be observed as the longest transmission. Although the useful data remains the same, \textit{i.e.}, $24$~bytes, the $ToA$ increases to $267.26$~ms due to the overhead introduced by the LoRaWAN stack, which typically adds about $12$~bytes of additional data, at the best. 

\begin{figure}[htpb]
    \begin{center}
        \includegraphics[keepaspectratio, scale=.5]{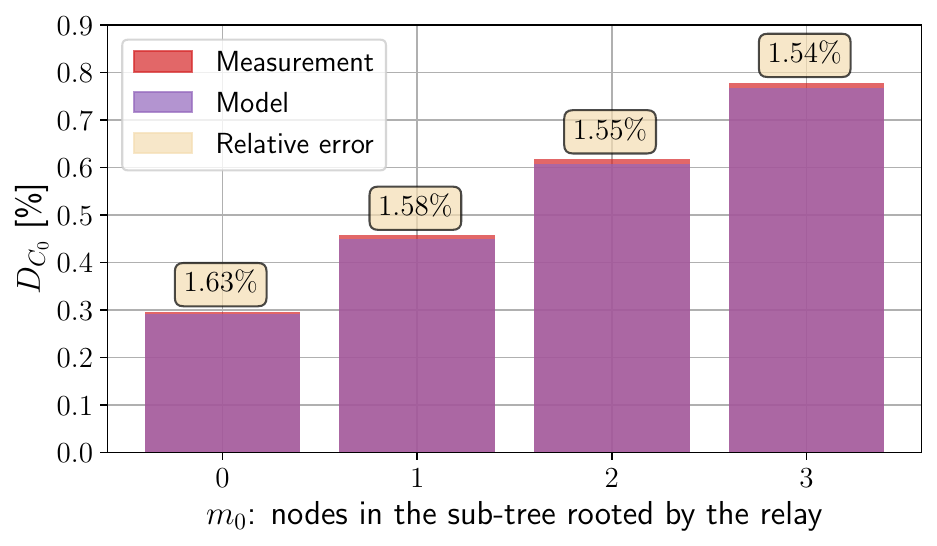}
        \caption{Measurement of the relay node duty cycle, $D_{C_0}$, together with the corresponding calculations obtained with \eqref{eq:dc_vs_mi} and the relative error between them for a star topology. For the measurements, an application period $T_{app}~=~4T_F$ was employed, jointly with $T_F=58.5$~s and $SF=9$.}
        \label{fig:measurements_dutycycle}
    \end{center}
\end{figure}

Following the experiment shown in Fig.~\ref{fig:tx_on_radio}, we calculated the total $T_x$ time of the relay during the $T_{app}$ period for various scenarios with different values $m_i$, in order to measure the relay's duty cycle. The findings are depicted in Fig.~\ref{fig:measurements_dutycycle}, together with the $D_{C_0}$ estimation obtained with the model~\eqref{eq:dc_vs_mi} and the relative error between them. The data reflects a linear behaviour in $m_0$, as expected from~\eqref{eq:dc_vs_mi}, and the maximum relative error obtained between the model and the measurements was $1.63\%$, which is mainly due to the difference between the ideal $ToA$ of the LoRa frames and the time employed by the radio chip to effectively do the transmission, as the radio frequency (RF) hardware and software always exhibit small delays that can slightly affect the radio time.

\section{Concluding Remarks and Future Work}
\label{sec:conclusions}

In the context of the abundance of IoT applications, certain domains remain underexplored in terms of the potential benefits that IoT could offer them. In this research, our focus was centered on addressing the need for automated wireless monitoring in the underground mining sector. To this end, we have developed, implemented, and evaluated a multi-hop TDMA protocol specifically tailored for inexpensive LoRa devices. With the use of this low overhead protocol, a proven sub $40$ $\mu$s synchronization accuracy between parent-child pairs was achieved for various topologies. \reviwer{This, combined with future efforts aimed at conducting field tests in underground scenarios, could eventually establish its complete suitability for use in such extreme assets}.
Furthermore, we presented a theoretical analysis which deals with the relationships between the frame size, network's throughput, and the sampling period of applications executed on the devices. Moreover, we introduced a model to estimate the duty cycle of the different members of the multi-hop network and its empirical measurements for validation. 
Our future work \reviwer{also} includes the reduction of the baseline cost to keep the network synchronized, \reviwer{together with the implementation of a simulation framework which will allow us to explore how scalable the proposed protocol implementation and model are when the number of devices increases.}

\bibliographystyle{./IEEEtran}
\bibliography{./IEEEabrv,./IEEEexample}


\end{document}